\documentstyle[twocolumn,prl,aps]{revtex} 

\input epsf 

\begin{document} 
\title{String breaking in zero-temperature lattice QCD} 
\author{\em UKQCD Collaboration}
\author{P. Pennanen\thanks{E-mail: {\tt petrus@hip.fi}}} 
\address{Nordita, Blegdamsvej 17, DK-2100 Copenhagen \O, Denmark} 
 \author{C. Michael\thanks{\tt cmi@liv.ac.uk}}\address{Theoretical  
Physics Division, Dept. of Math. Sciences, University of Liverpool,  
Liverpool, UK.} 
\maketitle \begin{abstract}
 The crossing from a static quark-antiquark system to a
system of two static-light mesons when the separation of the static 
quarks is increased is calculated in  zero-temperature lattice QCD. 
The mixing of these two states is extracted from the lattice operators. 
We also discuss the breaking of an excited string of a hybrid meson. \\ 
PACS numbers: 12.38.Aw 12.38.Gc 13.25.-k 13.75.Lb \end{abstract}
\section{Introduction}

The breaking of a long flux tube between two static quarks into a 
quark-antiquark pair is one of the most fundamental phenomena in QCD. 
Because  of  its highly non-perturbative nature it has defied
analytical calculation, while its large scale, e.g. when compared to
the sizes of composite particles in the theory, has caused difficulties in 
standard nonperturbative methods. Thus string breaking has remained a widely
publicized feature of the strong interaction  that has never, apart
from rough models, been reproduced  from the theory. 

String breaking can occur  in hadronic decays of $Q\bar{Q}$ mesons 
and is especially relevant when this meson is lying close to a 
meson-antimeson ($Q\bar{q}\bar{Q}q$) threshold. For the heaviest quarks 
involved in these decays applying heavy quark effective theory is
a reasonable approximation.

Due to recent advances in both computational hardware and algorithms,
much interest in lattice QCD has been devoted  to attempts to observe
string breaking. The direct approach of trying to see the flattening in
the static  $Q\bar{Q}$ potential at large separation has been successful
only at  temperatures close to the critical one~\cite{aok:98,lae:98}.
The failure  of this Wilson loop method at zero temperature seems to be
mainly due to the poor overlap of the operator(s) with the
$Q\bar{q}\bar{Q}q$   state~\cite{phi:98b,ste:99}. In much more easily
calculable adjoint $SU(2)$ and $SU(2)$+Higgs models without fermions a 
variational approach with explicit inclusion of both the Wilson loop and
scalar bound state operators has worked well
\cite{phi:98b,mic:92b,ste:99}. In three-dimensional SU(2) with
staggered fermions an improved action approach has been claimed to
be successful with just Wilson loops~\cite{tro:98}.
 In QCD with  fermions effective operators for $Q\bar{q}\bar{Q}q $
systems are, however, hard to implement; part of the problem is the
exhausting computational effort  required to get sufficient statistics
for light quark propagators with conventional techniques for fermion
matrix inversion. 

A new technique of calculating estimates of light quark propagators 
using Monte Carlo techniques on pseudo-fermionic field 
configurations~\cite{div:96} with maximal variance
reduction~\cite{mic:98} has been found to be very useful for systems
including heavy quarks taken as static, such as single heavy-light 
mesons and baryons and also two heavy-light mesons~\cite{mic:99,pen:99}. The
application of this method to the string breaking problem seems
natural. 

Our previous work \cite{mic:99,pen:99} has concentrated on bound states
of two heavy-light mesons and mechanisms of their attraction for
various values of  light quark isospin $I_q$ and spin $S_q$. Here we
continue by  studying the $Q\bar{q}\bar{Q}q$ system for $I_q,S_q=(0,1)$
together with the $Q\bar{Q}$ system at distances around the string
breaking point  $r_b\approx 1.2$ fm, where the energies of the two
systems are equal  if there is no mixing. 

\section{Quantum numbers and hybrid mesons}

When a quark and an antiquark are created from the vacuum they should have the
$0^{++}$ quantum numbers of the vacuum.  A quark-antiquark pair  with
$J^{PC}=0^{++}$ has lowest orbital angular momentum $L=1$  and is in a
spin triplet. This so-called Quark Pair Creation or $^3P_0$ model
was combined with  a harmonic oscillator flux tube model by Isgur {\em
et al.}~\cite{isg:85b} to  describe local breaking (formation) of a flux
tube. In our calculation the symmetries of the static approximation for
the heavy quark  automatically lead to only the light quark spin
triplet being nonzero, which  can be seen from the Dirac spin structure
of the heavy-light diagram in Fig.~\ref{fdiag}.

String breaking can also occur for hybrid mesons where the gluon field
between two quarks is in an excited state.  Table~\ref{texcite} presents
the  couplings of some low-lying gluonic excitations to the quantum
numbers of the  resulting meson-antimeson in the static limit for the
heavy quarks.  In this limit CP and $J_z$, where $z$ is the interquark
axis, are conserved.  The table lists  the representations of these
symmetries (with $\Sigma,\Pi, \Delta$ corresponding to $J_z=0,\ 1,\ 2$
respectively and $g,\ u$ corresponding to CP=$\pm 1$). For the lowest-lying
excitation, which has  $\Pi_u$ symmetry with $J_z=1$, only non-zero
angular momenta $L+L'$ for the  resulting mesons $B_{L'}$, $B_{L}$ would
be allowed, as   $S_q$ has to be zero to generate a negative $CP$. 

Here, as for the other symmetries, $I_q=0$ as we need the same flavour
for the light quarks. The $I_q=1$ cases do not correspond to
spontaneous breaking of a string in vacuum but a  correlation of a
$Q\bar{Q}$+$q\bar{q}$ system with a $Q\bar{q}$+$\bar{Q}q$  system at
different time, i.e. the breaking of a string in the presence of a meson.
When the table is extended to nonzero $L$ for the heavy quark plus
antiquark (as opposed to $L>0$ for a single meson) the $P,C$ values get
multiplied with $(-1)^L$. The energy levels also change;  these retardation effects have been found to be
relatively small~\cite{jug:99}. 

Previously string breaking in hybrid mesons has been discussed from a 
phenomenological point of view using  an
extension of the approach of Isgur {\em et al.}, i.e. a nonrelativistic
 flux-tube model with decay operators from strong coupling limit of
lattice gauge theory and heavy quark expansion of QCD in Coulomb 
gauge~\cite{pag:98}.
From this model two selection rules were given, the first one agreeing 
with the $\Pi_u$ case in Table~\ref{texcite}; low-lying hybrids do
not decay into identical mesons, the predominant channel being one $s$
and one $p$-wave meson. 
The second rule prohibits decay of spin singlet
states into only spin singlets, which is not relevant to our
calculation as our heavy quark spin decouples. 

An important question for hybrid meson phenomenology is the nature of
the  lowest state for a given set of quantum numbers at a particular
heavy quark separation; a hybrid $Q\bar{Q}$ meson, a ground-state
$Q\bar{Q}$ meson with a $q\bar{q}$ meson or a system of two heavy-light
mesons.  It is also useful to know the strength of mixing between these
states.  This information can be obtained, in principle,  from lattice
calculations and used to  decide what sort of bound states are most
likely to exist and what their decays will be (see also Ref.~\cite{mic:99b}).

\section{Lattice calculation}

We use SU(3) lattice QCD on a $16^3\times 24$ lattice with the Wilson 
gauge action and the  Sheikholeslami-Wohlert quark action with a
nonperturbative ``clover  coefficient'' $c_{SW}=1.76$ and $\beta=5.2$
with two degenerate flavours of both valence and sea quarks. The 
measurements were performed on  20 gauge configurations.  The gauge
configurations are the same as in Ref.~\cite{all:98} for 
$\kappa=0.1395$. With these  parameters we get a lattice spacing
$a\approx  0.14$ fm and meson mass ratio $M_{PS}/M_V=0.72$.

Estimators of propagators of quarks from  point $n$ to point $m$ can be
obtained from pseudofermion fields $\phi$. For each gauge configuration
a sample of the  pseudofermion fields is generated, and the propagators
are then obtained by Monte Carlo  integration~\cite{div:96}. Thus 
there is one Monte Carlo averaging for the gauge samples, and another
one for the pseudofermion samples for each gauge sample. In order to
reduce statistical variance of propagators a variance reduction method
similar to  multi-hit can be used~\cite{mic:98}; such a reduction is
essential in practice. Our variance reduction involves division of the
lattice in two regions, whose boundary is kept fixed while the
$\phi$-fields inside are replaced by their multi-hit averages. We use 24
pseudofermionic configurations per each gauge configuration. 

\begin{figure}[t]
\vspace*{-3.6cm}
\begin{center}
\hspace*{-2cm}\epsfxsize=400pt\epsfbox{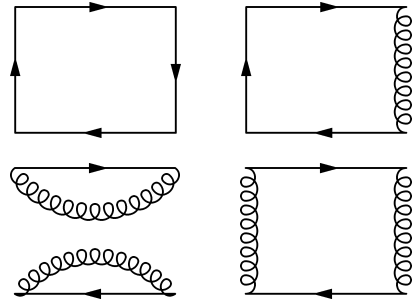}
\end{center}
\vspace{-13cm}
 \caption{Diagrams involved in the calculation; the Wilson loop $W$, the
heavy-light correlator $U$, the unconnected meson-antimeson $D$ and the
box diagram $B$.}
 \label{fdiag}
\end{figure}

Figure~\ref{fdiag} shows the diagrams involved in the calculation with
the time axis in the horizontal direction. The solid lines are heavy
quark propagators, which in the static approximation are just products
of gauge field variables. The wiggly lines are light quark
propagators, obtained essentially as a product of pseudofermionic
variables from each end which have to be in different variance
reduced regions~\cite{mic:98}.

In the large $T$ limit both the quark-antiquark and two-meson
operators should in  principle approach  $e^{-E_0(R) T}$ with $E_0(R)$
being the ground state of the system. In practice the Wilson loop has a
very small overlap with the two-meson state, which leads to great
practical difficulties in observing the flattening $E_0(R)\rightarrow
2M_{Q\bar{q}}$ from it at large $R$. The heavy-light term $U$
is necessary to obtain the correct ground state by explicitly including both
quark-antiquark and two-meson  states, and allows us to measure their
overlap, which is crucial for string breaking to happen. 

To estimate the ground (and excited) state energy of our observables 
we always use a variational basis formed from different degrees of
spatial  {\em fuzzing}  of the operators. This allows the use of
moderate values of $T$ instead of the infinite time limit to reduce
excited state contributions. The resulting correlation
matrix $C(R,T)$ is then diagonalised to get the eigenenergies. 

Due to the variance reduction method  dividing the lattice in two
halves in the time direction, the box diagram and the heavy-light
correlator in Fig.~\ref{fdiag}  have to be turned  ``sideways'' on the
lattice; i.e.,  the time axis in the diagrams is taken to be one of the
spatial axes to keep the light quark propagators going from one
variance reduced volume to another. This induces technical
complications that greatly increase the memory and  CPU demands of the
measurement program.

 For two flavours the $I_q=0$ wavefunction is of the form
$(u\bar{u}+d\bar{d})/\sqrt{2}$, which gives factors of $1,\ \sqrt{2},\ 
2$ for $D,\ U,\ B$ respectively. 
 For light quark spin we get the triplet states as in
Ref.~\cite{mic:99}.

In this first study we concentrate on the ground state breaking, i.e.
the  first row of Table~\ref{texcite}.  Investigation of the hybrid
meson breaking requires diagrams not  included in Fig.~\ref{fdiag},
which involve the hybrid $Q\bar{Q}$  and $Q\bar{Q}+\bar{q}q$ operators.
We estimate that for the $\Pi_u$ excited state the excited string
breaking  happens in the same distance range as for the ground state
due to the  non-zero momenta of the resulting mesons (masses taken from
 Ref.~\cite{mic:98}), which makes it harder to obtain sufficient
accuracy as the  spatial operators for excitations involve subtractions
rather than sums of  lattice paths. 

\section{Results}

\subsection{Variational approach}

A full variational matrix involving the Wilson loop, heavy-light
correlator and the $Q\bar{q}\bar{Q}q$ correlators gives the ground and
excited state energies and corresponding operator overlaps  as a
function of heavy quark separation, in analogue to the approach of 
Refs.~\cite{mic:92b,ste:99} for the adjoint string breaking and
Refs.~\cite{phi:98b} for the SU(2)+Higgs model. We use  a local light
quark creation (annihilation) operator and an extended  version where a
fuzzed path of link variables with length two separates  the operator
from the heavy quark line. For the link variables involved  in the
$Q\bar{Q}$ operators we have two fuzzing levels. The two  $Q\bar{Q}$ and
three $Q\bar{q}\bar{Q}q$ basis states then give a $5\times 5$
correlation matrix $C(R,T)$ that can be diagonalised.   However, for our 
present statistics the full matrix gives a reasonable
signal only for $r<r_b$. 

\begin{figure}[t]
\begin{center}
\epsfxsize=230pt\epsfbox{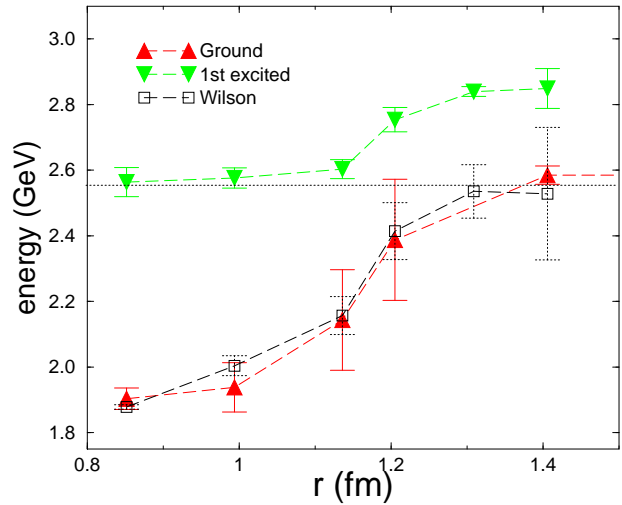}
\caption{Ground and excited state from a variational calculation including
$Q\bar{Q}$ and $Q\bar{q}\bar{Q}q$ operators. The highest fuzzed basis state
for both is used here. The ground state of the Wilson loop and 
$2m_{Q\bar{q}}$ are also shown. \label{fvari}}
\end{center}
\end{figure}

In Figure~\ref{fvari} the results from a calculation using just the most
fuzzed basis states for both $Q\bar{Q}$ and $Q\bar{q}\bar{Q}q$  (a
$2\times 2$ matrix) are shown. At $r_b$ we would expect the ground and
excited state energies to be separated by twice the mixing coefficient
$x$ (see below). We observe a larger separation which is presumably due
to our statistics not being sufficient to give accurate plateaus for the
energies. 

Although this full variational approach is in principle the most direct
way to study string breaking, we find that it is possible to focus on string
breaking explicitly, as we now discuss. 

\vspace*{-0.9cm}

\subsection{Mixing matrix element}

In full QCD, there is mixing of energy levels  between states  coupling
to Wilson lines (flux tube) and $Q\bar{q}\bar{Q}q$ states. To get  the
mixing matrix element the correlation between a Wilson line and a
$Q\bar{q}\bar{Q}q$ operator has to be considered. In order to study the
operator mixing from this heavy-light correlator one needs to use
results (energies and couplings) from both diagonal operators
separately: thus  from the Wilson loop  (with ground state contribution
given by $W(T)=w^2 \exp[-V(R)T]$)  and the unconnected $Q\bar{q}\bar{Q}q$
correlator (eg. $D(T)=d^2 \exp[-M(R)T]$ from the ground state)
where we use a variational basis to suppress excited states.

The ground state contribution to the heavy-light correlator can
then be written as
 \begin{equation}
U(T)  =  x(R) \sum_{t=0}^T w e^{-V(R)t}e^{-M(R)(T-t)} d+O(x^3)
 \label{ehlcq}
 \end{equation}
 In the quenched case the contributions from fermion loops inside the
correlator are absent, removing the $O(x^3)$ terms in Eq.~\ref{ehlcq}.
The box term is expressed in the same manner as
 \begin{eqnarray}
B(T) & = &  x^2(R) \sum_{t_1=0}^T  \sum_{t_2 \ge t_1}^T
d e^{-V(R)t_1} e^{-M(R)(t_2-t_1)}
\nonumber\\
 & & \times \ e^{-V(R)(T-t_2)} d \ +\ O(x^4)  \label{ebcq} 
 \end{eqnarray}

The operator mixing coefficient $x$ for the $Q\bar{Q}$ and
$Q\bar{q}\bar{Q}q $ states can be  extracted from these expressions.
Near the  string breaking point (where $V(R) =M(R)$),  in the infinite
time limit, only  the ground state contributions survive.
 We use
 \begin{eqnarray}
 x & = & \frac{U(T)}{\sqrt{W(T) D(T)}}\frac{f^{T/2}}{1+\ldots +f^T} +
O(x^3) \label{ex1} \\
  & = &  \sqrt{\frac{B(T)}{D(T)}}
\frac{f^{T/2}}{\sqrt{1+\ldots+(T+1)f^T}}  +O(x^2) \label{ex2}
 \end{eqnarray}
 The factors of $f\equiv\exp(V(R)-M(R))$ account for departures  from
the string breaking point.                                   
    
  In the quenched case there is no mixing between the energy levels of
the quark-antiquark and two-meson systems, and $x$ can  be
extracted using  Eqs.~\ref{ex1},\ref{ex2}.  As $x<<1$ the
non-leading terms in the expressions for $x$ are small and we may use
also these formulas with our unquenched data - with a resulting decrease
in errors compared  to the full variational study of the preceeding
subsection.

 Our assumption about neglecting excited state contributions can be
tested by obtaining consistent results for $x$ from both relations for
several  $T$ values. To improve further on our estimate of $x$ we
diagonalise separately  $W,\ D$ and $B$ to enhance the ground state
contributions and use the first two diagonalisations to extract the ground state  of
$U$. Our results with bootstrap errors can be seen in table~\ref{tx}.
Assuming constant $x$ for $0.99 \ {\rm fm} \le r \le 1.31$ fm gives us a best
estimate of  $x/a=0.033(6)/a=46(8)$ MeV. This is about half of the value
of $x=100$ MeV obtained using a strong coupling mixing model and the
experimental $\Upsilon(4S)$ decay  rate~\cite{dru:98a}.    

 Our analysis shows that the string breaking matrix element is
small but non-zero. We are able, however, to find two independent ways
to estimate it (using all four diagrams in Fig.~1) and we
obtain $x=46(8)$ MeV with light quarks that are  around the strange
quark mass.  This is the first non-perturbative  determination from QCD
of the string breaking matrix element. Because of its  small value,
direct observation of string breaking from the spectrum is
difficult to achieve. 
				
{\bf Acknowledgement}

We thank A.M. Green and K. Rummukainen for discussions. Some of the
calculations, consuming 70 GB of disk and $2\times 10^{16}$ FLOPs, were 
performed with the excellent resources provided by the CSC in Espoo, Finland.

\vspace*{-0.5cm}

\newcommand{\href}[2]{#2}\begingroup\raggedright\endgroup

\begin{table}[tb]
\begin{center}
\begin{tabular}{ll|cc|ccc}
$J_z$ & $CP$ & Gluon field symmetry & $\Sigma_{1g}$+$q\bar{q}$ state & $I_q$ & $S_q$ & $L+L'$  \\ \hline
0   & $+$  & $\Sigma_{1g}$             & $\omega$ & 0 & 1 & 0  \\
1   & $-$  & $\Pi_u$                & $h$      &   & 0   & 1  \\
0   & $-$  & $\Sigma_{1u}$             & $\eta$   &   & 0   & 0  \\
1   & $+$  & $\Pi_g$                & $\omega$ &   & 1   & 0  \\
2   & $+$  & $\Delta_{g}$             & $f_2$    &   & 1   & 1 \\
2   & $-$  & $\Delta_{u}$             & $\eta_2$  &   & 0   & 2 \\
 \hline
0   & $-$  & --                   & $\pi$    & 1 & 0   & 0  \\
0   & $+$  & --                   & $\rho$   &   & 1   & 0 
\end{tabular}
\caption{Relation of gluonic excitations of a hybrid meson to ground-state 
properties of the meson pair resulting from string breaking, in the 
static limit for the heavy quarks. The last three columns refer to the 
quantum numbers of the light quarks in the meson-antimeson system. \label{texcite}}
\end{center}
\end{table} 

\vspace*{-1.1cm}

\begin{table}[hb]
\begin{center}
\begin{tabular}{l|cc}
$r$ (fm) & Eq.~\ref{ex1} without $O(x^3)$ & Eq.~\ref{ex2} without $O(x^2)$\\ \hline
0.85    &  0.040(3) & 0.081(8) \\
0.99    &  0.025(4) & 0.045(5) \\
1.14    &  0.025(3) & \\
1.31    & 0.031(15) & 0.040(25) \\
\end{tabular}

\caption{Operator mixing $x$ extracted using a variational approach with two 
different formulas. The values are taken at $T=4$ and have bootstrap errors.
Physical dimensions are obtained by multiplying with $a^{-1}=1.39$ GeV \label{tx}}
\end{center}
\end{table} 
\vspace*{-1cm}

\end{document}